\documentclass[aps,prd,twocolumn,showpacs,showkeys,amsmath,amssymb]{revtex4}
\usepackage{graphicx,color}
\usepackage{dcolumn}
\usepackage{bm}
\begin{document}
\author{Wei Wei}
\author{Xiang Liu}
\author{Shi-Lin Zhu}
\affiliation{Department of Physics, Peking University, Beijing
100871, China}

\title{D Wave Heavy Mesons} \vspace{1.cm}

\begin{abstract}
We first extract the binding energy $\bar \Lambda$ and decay
constants of the D wave heavy meson doublets $(1^{-},2^{-})$ and
$(2^{-},3^{-})$ with QCD sum rule in the leading order of heavy
quark effective theory. Then we study their pionic $(\pi, K, \eta)$
couplings using the light cone sum rule, from which the parameter
$\bar \Lambda$ can also be extracted. We then calculate the pionic
decay widths of the strange/non-strange D wave heavy $D/B$ mesons
and discuss the possible candidates for the D wave charm-strange
mesons. Further experimental information, such as the ratio between
$D_s\eta$ and $DK$ modes, will be very useful to distinguish various
assignments for $D_{sJ}(2860, 2715)$.
\end{abstract}


\maketitle
\pagenumbering{arabic}

\section{Introduction}

Recently BarBar reported two new $D_{s}$ states, $D_{sJ}(2860)$
and $D_{sJ}(2690)$ in the $DK$ channel. Their widths are
$\Gamma=48\pm7\pm10 $ MeV and $\Gamma=112\pm7\pm36 $ MeV
respectively \cite{babar}. For $D_{sJ}(2860)$ the significance of
signal is $5\sigma$ in the $D^0K^+$ channel and 2.8 $\sigma$ in
the $D^+K_s^0$ channel. Belle observed another state
$D_{sJ}(2715)$ with $J^P=1^{-}$ in $B^+\rightarrow \bar{D^0}D_{sJ}
\rightarrow\bar{D^0}D^0K^+$ \cite{belle}. Its width is
$\Gamma=115\pm20$ MeV. No $D^{*}K$ or $D_s\eta$ mode has been
detected for all of them.

The $J^P$ of $D_{sJ}(2860)$ and $D_{sJ}(2690)$ can be $0^+,1^-,
2^+,3^-,\cdots$ since they decay to two pseduscalar mesons.
$D_{sJ}(2860)$ was proposed as the first radial excitation of
$D_{sJ}(2317)$ based on a coupled channel model \cite{rupp} or an
improved potential model \cite{close}. Colangelo et al considered
$D_{sJ}(2860)$ as the D wave $3^{-}$ state \cite{colangelo}. The
mass of $D_{sJ}(2715)$ or $D_{sJ}(2690)$ is consistent with the
potential model prediction of  the $c\bar{s}$ radially excited
$2^3S_1$ state \cite{isgur,close}. Based on chiral symmetry
consideration, a D wave $1^-$ state with mass $M=2720 $ MeV is also
predicted if the $D_{sJ}(2536)$ is taken as the P wave $1^+$ state
\cite{nowak}. The strong decay widths for these states are discussed
using the $^{3}P_{0}$ model in \cite{bozhang}.

The heavy quark effective theory (HQET) provides a systematic
expansion in terms of $1/ m_Q$ for hadrons containing a single heavy
quark, where $m_Q$ is the heavy quark mass \cite{grinstein}. In HQET
the heavy mesons can be grouped into doublets with definite
$j_{\ell}^P$ since the angular monument of light components
$j_{\ell}$ is a good quantum number in the $m_Q\to\infty$ limit.
They are $\frac{1}{2}^-$ doublet $(0^-, 1^-)$ with orbital angular
monument $L=0$, $\frac{1}{2}^+$ doublet $(0^+,1^+)$ and
$\frac{3}{2}^+$ doublet $(1^+,2^+)$ with $L=1$. For $L=2$ there are
$(1^{-},2^{-})$ and $(2^{-},3^{-})$ doublets with
$j_{\ell}^P=\frac{3}{2}^-$and $\frac{5}{2}^-$ respectively. The
states with the same $J^P$, such as the two $1^-$ and two $1^+$
states, can be distinguished in the $m_Q\to\infty$ limit, which is
one of the advantage of working in HQET. The D wave heavy mesons
(($B_1^{*'} ,B_2^{*}$), ($B_2^{*'}, B_3$)) were considered in the
quark model \cite{quark model}. The semileptonic decay of $B$ meson
to the D wave doublets was calculated using three point QCD sum rule
\cite{colangelo2}. The decay property of the heavy mesons till $L=2$
was calculated using the $^{3}P_{0}$ model in \cite{close2}.

Light cone QCD sum rule (LCQSR) has proven very useful in extracting
the hadronic form factors and coupling constants in the past decade
\cite{light cone}. Unlike the traditional SVZ sum rule \cite{svz},
it was based on the twist expansion on the light cone. The strong
couplings and semileptonic decay form factors of the low lying heavy
mesons have been calculated using this method both in full QCD and
in HQET \cite{lc}.

In this paper we first extract the mass parameters and decay
constants of D wave doublets in section \ref{mass}. Then we study
the strong couplings of the D wave heavy doublets with light
pseduscalar mesons $\pi$, $K$ and $\eta$ in section \ref{lcqsr}. We
work in the framework of LCQSR in the leading order of HQET. We
present our numerical analysis in section \ref{numerical}. In
section \ref{width} we calculate the strong decay widths to light
hadrons and discuss the possible D wave charm-strange heavy meson
candidates. The results are summarized in section \ref{summary}.

\section{Two-point QCD sum rules}\label{mass}

The proper interpolating currents
$J_{j,P,j_{\ell}}^{\alpha_1\cdots\alpha_j}$ for the states with the
quantum number $j$, $P$, $j_{\ell}$ in HQET were given in
\cite{huang}, with $j$ the total spin of the heavy meson, $P$ the
parity and $j_{\ell}$ the angular momentum of light components. In
the $m_Q\to\infty$ limit, the currents satisfy the following
conditions
\begin{eqnarray}
\label{decay} &&\langle
0|J_{j,P,j_{\ell}}^{\alpha_1\cdots\alpha_j}(0)|j',P',j_{\ell}^{'}\rangle=
f_{Pj_l}\delta_{jj'}
\delta_{PP'}\delta_{j_{\ell}j_{\ell}^{'}}\eta^{\alpha_1\cdots\alpha_j}\;,\nonumber\\
\label{corr}&&i\:\langle 0|T\left
(J_{j,P,j_{\ell}}^{\alpha_1\cdots\alpha_j}(x)J_{j',P',j_{\ell}'}^{\dag
\beta_1\cdots\beta_{j'}}(0)\right )|0\rangle=
\delta_{jj'}\delta_{PP'}\delta_{j_{\ell}j_{\ell}'}\nonumber\\
&&\times\:(-1)^j\:{\cal S}\:g_t^{\alpha_1\beta_1}\cdots
g_t^{\alpha_j\beta_j} \int \,dt\delta(x-vt)\:\Pi_{P,j_{\ell}}(x)\;,\nonumber\\
\end{eqnarray}
where $\eta^{\alpha_1\cdots\alpha_j}$ is the polarization tensor for
the spin $j$ state. $v$ denotes the velocity of the heavy quark. The
transverse metric tensor
$g_t^{\alpha\beta}=g^{\alpha\beta}-v^{\alpha}v^{\beta}$. ${\cal S}$
denotes symmetrizing the indices and subtracting the trace terms
separately in the sets $(\alpha_1\cdots\alpha_j)$ and
$(\beta_1\cdots\beta_{j})$. $f_{P,j_{\ell}}$ is a constant.
$\Pi_{P,j_{\ell}}$ is a function of $x$. Both of them depend only on
$P$ and $ j_{\ell}$.

The interpolating currents are \cite{huang}
\begin{eqnarray}
\label{curr1} &&J^{\dag\alpha}_{1,-,{3\over
2}}=\sqrt{\frac{3}{4}}\:\bar h_v(-i)\left(
{\cal D}_t^{\alpha}-\frac{1}{3}\gamma_t^{\alpha}{\cal {D}\!\!\!\slash}_t\right)q\;,\\
\label{curr2} &&J^{\dag\alpha_1,\alpha_2}_{2,-,{3\over
2}}=\sqrt{\frac{1}{2}}\:T^{\alpha_1,\alpha_2;\;\beta_1,\beta_2}\bar
h_v (-i)\nonumber\\
&&\qquad\quad\times\:\left({\cal D}_{t \beta_1}{\cal D}_{t
\beta_2}-\frac{2}{5}{\cal D}_{t \beta_1} \gamma_{t \beta_2}
{\cal D\!\!\!\slash}_t\right)q\;,\\
\label{curr3} &&J^{\dag\alpha_1,\alpha_2}_{2,-,{5\over
2}}=-\sqrt{\frac{5}{6}}\:T^{\alpha_1,\alpha_2;\;\beta_1,\beta_2}\bar
h_v \gamma^5 \nonumber\\
&&\qquad\quad\times\:\left({\cal D}_{t \beta_1}{\cal D}_{t
\beta_2}-\frac{2}{5}{\cal D}_{t\beta_1} \gamma_{t\beta_2}
{\cal D\!\!\!\slash}_t\right)q\;,\\
\label{curr4}&&J^{\dag\alpha,\beta,\lambda}_{3,-,{5\over
2}}=-\sqrt{\frac{1}{2}}\:T^{\alpha,\beta,\lambda;\;\mu,\nu,\sigma}\bar
h_v
\gamma_{t\mu}{\cal D}_{t\nu}{\cal D }_{t\sigma} q\;,\\
\label{curr5} &&J^{\dag\alpha}_{1,-,{1\over
2}}=\sqrt{\frac{1}{2}}\:\bar h_v\gamma_t^{\alpha} q\;,\hspace{0.4cm}
J^{\dag}_{0,-,{1\over
2}}=\sqrt{\frac{1}{2}}\:\bar h_v\gamma_5q\;,\\
\label{curr6} &&J^{\dag\alpha}_{1,+,{1\over
2}}={\sqrt{\frac{1}{2}}}\:\bar h_v\gamma^5\gamma^{\alpha}_tq\;,\\
\label{curr7} &&J^{\dag\alpha}_{1,+,{3\over
2}}=\sqrt{\frac{3}{4}}\:\bar h_v\gamma^5(-i)\left(
{\cal D}_t^{\alpha}-\frac{1}{3}\gamma_t^{\alpha}{\cal D\!\!\!\slash}_t\right)q\;,\\
\label{curr8} &&J^{\dag\alpha_1,\alpha_2}_{2,+,{3\over
2}}=\sqrt{\frac{1}{2}}\:\bar h_v
\frac{(-i)}{2}\nonumber\\
&&\qquad\quad\times\:\left(\gamma_t^{\alpha_1}{\cal D}_t^{\alpha_2}+
\gamma_t^{\alpha_2}{\cal D}_t^{\alpha_1}-{2\over
3}g_t^{\alpha_1\alpha_2} {\cal D\!\!\!\slash}_t\right)q\;,
\end{eqnarray}
where $h_v$ is the heavy quark field in HQET and $\gamma_{t\mu}=\gamma_%
\mu-v_\mu {v}\!\!\!\slash$. The definitions of
$T^{\alpha,\beta;\;\mu,\nu}$ and
$T^{\alpha,\beta,\lambda;\;\mu,\nu,\sigma}$ are given in Appendix
\ref{appendix1}.

We first study the two point sum rules for the ($1^-, 2^-$) and
($2^-, 3^-$) doublets. We consider the following correlation
functions: {\small
\begin{eqnarray}
&&\label{correlator 1} i \int d^4x\: e^{ikx}\langle
\pi|T(J^{\alpha}_{1,-,{3\over 2}}(x)J^{\beta}_{1,-,{3\over
2}})|0\rangle=-g_t^{\alpha\beta}\Pi_{-,{3\over
2}}(\omega)\;, \nonumber\\
~\\
 && \label{correlator 2}i \int d^4 x e^{ikx}\langle
\pi|T(J^{\alpha_1\alpha_2}_{2,-,{5\over
2}}(x)J^{\beta_1\beta_2}_{2,-,{5\over
2}})|0\rangle=\frac{1}{2}(g_t^{\alpha_1\beta_1}g_t^{\alpha_2\beta_2}
\nonumber\\
&&\qquad
+g_t^{\alpha_1\beta_2}g_t^{\alpha_2\beta_1}-\frac{2}{3}g_t^{\alpha_1\alpha_2}g_t^{\beta_1\beta_2})\Pi_{-,{5\over
2}}(\omega)\;,
\end{eqnarray}}
where $\omega=2v\cdot k$.

At the hadron level,
\begin{equation}
\Pi_{P,j_{\:l}}=\frac{f^2_{P,j_l}}{2\bar{\Lambda}_{P,j_l}-\omega}\nonumber
+\cdots\;.\nonumber
\end{equation}
At the quark-gluon level it can be calculated with the leading order
lagrangian in HQET. Invoking quark-hadron duality and making the
Borel transformation, we get the following sum rules from eqs.
(\ref{correlator 1}) and (\ref{correlator 2})
\begin{eqnarray}
&&f_{-,{3\over2}}^2\exp\Big[{-{2\bar\Lambda_{-,{3\over2}}\over
T}}\Big]\nonumber\\&&={1\over
2^6\pi^2}\int_0^{\omega_c}\omega^4e^{-\omega/{T}}d\omega
+\frac{1}{16}\:m_0^2\:\langle\bar qq\rangle -{1\over
2^5}\langle{\alpha_s\over\pi}G^2\rangle T,\nonumber\\\label{form1}
&&f_{-,{5\over2}}^2\exp\Big[{-{2\bar\Lambda_{-,{5\over2}}\over
T}}\Big]\nonumber\\&&={1\over 2^7\cdot
5\pi^2}\int_0^{\omega_c}\omega^6e^{-\omega/{T}}d\omega+{1\over
120}\langle{\alpha_s\over\pi}G^2\rangle T^3\;\label{form2}.
\end{eqnarray}
Here $m_0^2\,\langle\bar qq\rangle=\langle\bar
qg\sigma_{\mu\nu}G^{\mu\nu}q\rangle$. Only terms of the lowest order
in $\alpha_s$ and operators with dimension less than six have been
included. For the ${5\over 2}^-$ doublet there is no mixing
condensate due to the higher derivation.

We use the following values for the QCD parameters: $\langle\bar
qq\rangle=-(0.24 ~\mbox{GeV})^3$, $\langle\alpha_s GG\rangle=0.038
~\mbox{GeV}^4$, $ m_0^2=0.8 ~\mbox{GeV}^2$.

Requiring that the high-order power corrections is less than
$30\%$ of the perturbation term without the cutoff $\omega_c$ and
the contribution of the pole term is larger than $35\%$ of the
continuum contribution given by the perturbation integral in the
region $\omega
> \omega_c$, we arrive at the stability region of the sum rules
$\omega_c=3.2-3.6$ GeV, $T=0.8-1.0$ GeV.

\begin{figure}[hbt]
\begin{center}
\scalebox{0.8}{\includegraphics{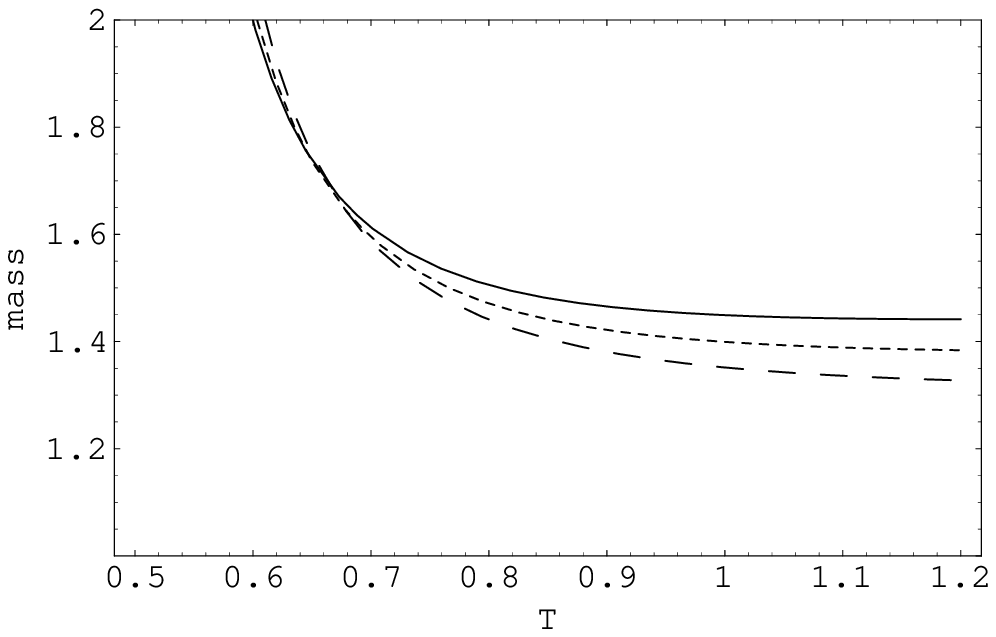}}
\end{center}
\caption{\label{fig1} The variation of the mass parameter
$\bar\Lambda_{-,{3\over2}}$ with the threshold $\omega_c$ (in unit
of $\mbox{GeV}$) for the ${3\over 2}^-$ doublet. The long-dashed,
short dashed and solid curves correspond to $\omega_c$=3.2, 3.4, 3.6
GeV respectively.}
\end{figure}

\begin{figure}[hbt]
\begin{center}
\scalebox{0.8}{\includegraphics{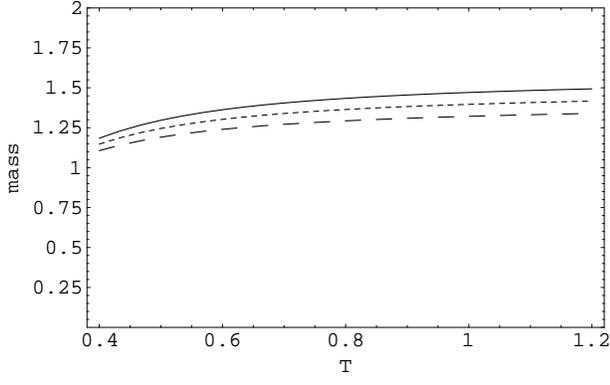}}
\end{center}
\caption{\label{fig2}The variation of the mass parameter
$\bar\Lambda_{-,{5\over2}}$ with the threshold $\omega_c$ (in unit
of $\mbox{GeV}$) for the ${5\over 2}^-$ doublet. The long-dashed,
short dashed and solid curves correspond to $\omega_c$=3.2, 3.4, 3.6
GeV respectively.}
\end{figure}

The results for $\bar\Lambda$'s are
\begin{eqnarray}
\label{result1}
\bar\Lambda_{-,{3\over2}}&=&1.42 \pm 0.08 ~~\mbox{GeV}\;,\\
\bar\Lambda_{-,{5\over2}}&=&1.38 \pm 0.09 ~~\mbox{GeV}\;.
\end{eqnarray}
The errors are due to the variation of $T$ and the uncertainty in
$\omega_c$. In Fig. \ref{fig1} and \ref{fig2}, we show the
variations of masses with $T$ for different $\omega_c$.


The masses of D wave mesons in the quark model are around $2.8~
\mbox{GeV}$ for $D$ meson and $6 ~\mbox{GeV}$ for $B$ meson
\cite{quark model}. $1/m_Q$ correction may be quite important for D
wave heavy mesons, which will be investigated in a subsequent work.

In the following sections we also need the values of $f$'s:
\begin{eqnarray}
 f_{-, {3\over2}}&=&0.39\pm 0.03 ~\mbox{GeV}^{5/2}\;,\\
 f_{-, {5\over2}}&=&0.33\pm 0.04 ~\mbox{GeV}^{7/2}\;.
\end{eqnarray}

\section{ Sum rules for decay amplitudes}\label{lcqsr}

Now let us consider the strong couplings of D wave doublets ${3\over
2}^-$ and ${5 \over 2}^-$ with light hadrons. For the light quark
being a $u$ (or $d$) quark, the D wave heavy mesons decay to pion,
while for the light quark being a strange quark, it can decay either
to $BK$ or $B_s\eta$. In the following we will denote the light
hadron as pion and discuss all three cases. The strong decay
amplitudes for D wave $1^-$ and $3^-$ states to ground doublet
$(0^-,1^-)$ are
\begin{eqnarray}
&&M(B_1^{*'}\rightarrow B\pi)=I \epsilon^{\mu}q_{t\mu}g(B_1^{*'}B)\;,\nonumber\\
&&M(B_1^{*'}\rightarrow B^{*}\pi)
=I\:i\epsilon^{\mu\nu\rho\sigma}\epsilon_{\mu}\epsilon^{'*}_{\nu}v_{\rho}q_{t\sigma}g(B_1^{*'}B^{*})\;,\nonumber\\
&&M(B_3\rightarrow B\pi)
=I\:\epsilon^{\alpha\beta\lambda}(q_{t\alpha}q_{t\beta}q_{t\lambda}-
 \frac{1}{6}q^2_t(g_{t\alpha\beta}q_{t\lambda} \nonumber\\
 &&\qquad\qquad\qquad\qquad
 + g_{t\alpha\lambda}q_{t\beta}+
 {4\over 3}g_{t\beta\lambda}q_{t\alpha} ))g(B_3B)\;,\nonumber\\
&&M(B_3\rightarrow B^{*}\pi)
=I\:i\epsilon^{\mu\nu\sigma\alpha}\:\epsilon_{\alpha\beta\lambda}\:\epsilon^{*}_{\mu}\:v_{\sigma}\;,\nonumber\\
 &&\qquad\qquad\qquad\qquad \times
\bigg[q_{t\nu}\:q_{t}^{\beta}\:q_{t}^{\lambda}-\frac{1}{6}q^2_t\bigg(g_{t\nu}^{\beta}q_t^{\lambda}
+ g_{t\nu}^{\lambda}q_t^{\beta}\nonumber\\
 &&\qquad\qquad\qquad\qquad +
 {4\over 3}g_t^{\beta\lambda}q_{t\nu}\bigg)\bigg]\:g(B_3B^{*})\;,\label{amp}
\end{eqnarray}
where  $\epsilon^{\alpha\beta\lambda}$, $\epsilon^{\mu}$, and
$\epsilon^{'}_{\nu}$ are  polarizations of the D wave $3^-$, $1^-$
states and ground $1^-$ state respectively. $I=1,
\frac{1}{\sqrt{2}}$ for charged and neutral pion mesons
respectively, while for the $K$ and $\eta$ mesons it equals one.
$g(B_1^{*'}B)$ etc are the coupling constants in HQET and are
related to those in full QCD by
\begin{equation}
g^{\mbox{\tiny{full
QCD}}}(B_1^{*'}B)=\sqrt{m_{B_1^{*'}}m_{B}}\;g^{\mbox{\tiny{HQET}}}(B_1^{*'}B)\;.
\end{equation}

Because of the heavy quark symmetry, the coupling constants in eq.
(\ref{amp}) satisfy
\begin{eqnarray}
g(B_1^{*'}B)&=&g(B_1^{*'}B^{*})\;,\nonumber\\
g(B_3^{*}B)&=&g(B_3^{*}B^{*})\;.
\end{eqnarray}

In order to derive the sum rules for the coupling constants we
consider the correlators
\begin{widetext}
\begin{eqnarray}
&& \int d^4x\;e^{ik\cdot
x}\langle\pi(q)|T\left(J^{\alpha}_{1,-,\frac{3}{2}}(x)
 J^{\dagger}_{0,-,\frac{1}{2}}(0)\right)|0\rangle =
         q_t^{\alpha}I\;G_{1}(\omega,\omega')\;,\label{c1}\\
&& \int d^4x\;e^{ik\cdot
x}\langle\pi(q)|T\left(J^{\alpha}_{1,-,\frac{3}{2}}(x)
 J^{\dagger \beta}_{1,+,\frac{1}{2}}(0)\right)|0\rangle =
         (q_t^{\alpha}q_t^{\beta}-\frac{1}{3}g_t^{\alpha\beta}q_t^2)I\;G_{2}(\omega,\omega')\;,\label{c2}\\
&& \int d^4x\;e^{ik\cdot
x}\langle\pi(q)|T\left(J^{\beta}_{1,+,\frac{3}{2}}(x)
 J^{\dagger\alpha}_{1,-,\frac{3}{2}}(0)\right)|0\rangle =
         (q_t^{\alpha}q_t^{\beta}-\frac{1}{3}g_t^{\alpha\beta}q_t^2)I\;G_{3}^d(\omega,\omega')
         +g_t^{\alpha\beta}I G_{3}^s\;,\label{c3}\\
&& \int d^4x\;e^{ik\cdot
x}\langle\pi(q)|T\left(J^{\alpha}_{1,-,\frac{3}{2}}(x)
 J^{\dagger\alpha_1\alpha_2}_{2,-,\frac{3}{2}}(0)\right)|0\rangle
 =\Big[\frac{1}{2}(g_t^{\alpha\alpha_1}q_t^{\alpha_2}+g_t^{\alpha\alpha_2}q_t^{\alpha_1})
-\frac{1}{3}g_t^{\alpha_1\alpha_2}q_t^{\alpha}\Big]I \;G_{4}^p(\omega,\omega')\nonumber\\
&&  \qquad \qquad \qquad \qquad \qquad \qquad\qquad \qquad +\Big[
q_t^{\alpha}q_t^{\alpha_1}q_t^{\alpha_2}-
 \frac{1}{6}q^2_t(g_t^{\alpha\alpha_1}q_t^{\alpha_2} +
 g_t^{\alpha\alpha_2}q_t^{\alpha_1} +
 {4\over 3}g_t^{\alpha_1\alpha_2}q_t^{\alpha} )\Big]I\;
G_{4}^f(\omega,\omega')\;,
\end{eqnarray}
for the $j_{\ell}^P=\frac{3}{2}^-$ doublet;
\begin{eqnarray}
&& \int d^4x\;e^{ik\cdot
x}\langle\pi(q)|T\left(J^{\alpha\beta\lambda}_{3,-,\frac{5}{2}}(x)
 J^{\dagger}_{0,-,\frac{1}{2}}(0)\right)|0\rangle =
\Big[q_t^{\alpha}q_t^{\beta}q_t^{\lambda}-
 \frac{1}{6}q^2_t(g_t^{\alpha\beta}q_t^{\lambda} +
 g_t^{\alpha\lambda}q_t^{\beta} +
 {4\over 3}g_t^{\beta\lambda}q_t^{\alpha}
 )\Big ]I\;G_{5}(\omega,\omega')\;,
\\
&& \int d^4x\;e^{ik\cdot
x}\langle\pi(q)|T\left(J^{\alpha\beta}_{2,-,\frac{5}{2}}(x)
 J^{\dagger }_{0,+,\frac{1}{2}}(0)\right)|0\rangle =
         (q_t^{\alpha}q_t^{\beta}-\frac{1}{3}g_t^{\alpha\beta}q_t^2)I\;G_{6}(\omega,\omega')\;,\label{c6} \end{eqnarray}
 \begin{eqnarray}
&&\int d^4x\;e^{ik\cdot
x}\langle\pi(q)|T\left(J^{\alpha\beta}_{2,-,\frac{5}{2}}(x)
 J^{\dagger \gamma}_{1,+,\frac{3}{2}}(0)\right)|0\rangle =
         \frac{1}{2}i(\epsilon^{\beta\gamma\mu\nu}q_t^{\alpha}+\epsilon^{\alpha\gamma\mu\nu}q_t^{\beta})
         q_{t\mu}v_{\nu}I \;G_{7}(\omega,\omega')\;,\label{c7}
\end{eqnarray}
\begin{eqnarray}
&&\int d^4x\;e^{ik\cdot
x}\langle\pi(q)|T\left(J^{\alpha\beta}_{2,-,\frac{5}{2}}(x)
 J^{\dagger\lambda}_{1,-,\frac{3}{2}}(0)\right)|0\rangle =\Big[\frac{1}{2}(g_t^{\alpha\lambda}q_t^{\beta} +
 g_t^{\beta\lambda}q_t^{\alpha})-\frac{1}{3}g_t^{\alpha\beta}q_t^{\lambda}\Big]I\;G^p_{8}(\omega,\omega')\nonumber\\
&&\qquad \qquad \qquad \qquad \qquad \qquad \qquad \qquad
\qquad\qquad +\Big[q_t^{\alpha}q_t^{\beta}q_t^{\lambda}-
 \frac{1}{6}q^2_t(g_t^{\alpha\lambda}q_t^{\beta} +
 g_t^{\beta\lambda}q_t^{\alpha} +
 {4\over 3}g_t^{\alpha\beta}q_t^{\lambda}
 )\Big] I\;G^f_{8}(\omega,\omega')\;,\label{c8}\\
&& \int d^4x\;e^{ik\cdot
x}\langle\pi(q)|T\left(J^{\alpha\beta}_{2,-,\frac{5}{2}}(x)
 J^{\dagger\mu\nu\sigma}_{3,-,\frac{5}{2}}(0)\right)|0\rangle
 =T^g I\; G_{9}^g(\omega,\omega')+T^{f} I\;G_{9}^{f
        }(\omega,\omega')+ T^{p1} I\; G_{9}^{p1}(\omega,\omega')\nonumber\\
 &&\qquad \qquad \qquad \qquad \qquad \qquad \qquad \qquad \qquad\qquad
 + T^{p2}I\; G_{9}^{p2}(\omega,\omega')\;,\label{c9}
\end{eqnarray}
\end{widetext}
for the $j_{\ell}^P=\frac{5}{2}^-$ doublet, where $k^{\prime}=k+q$,
$\omega=2v\cdot k$, $\omega^{\prime}=2v\cdot k^{\prime}$. Note that
the two P wave couplings between $({2,-,\frac{5}{2}})$ and
$({3,-,\frac{5}{2}})$ in eq. (\ref{c9}) are not independent and
satisfy the relation $g_9^{p2}=-\frac{1}{3}g_9^{p1}$.

First let us consider the function $G_1(\omega,\omega^{\prime})$ in
eq. (\ref{c1}). As a function of two variables, it has the following
pole terms from the double dispersion relation
\begin{eqnarray}
\label{pole} {f_{-,{1\over 2}}f_{-,{3\over 2}}g_1\over
(2\bar\Lambda_{-,{1\over 2}} -\omega')(2\bar\Lambda_{-,{3\over
2}}-\omega)}+{c\over 2\bar\Lambda_{-,{1\over 2}} -\omega'}+{c'\over
2\bar\Lambda_{-,{3\over 2}}-\omega}\;,~~\nonumber
\end{eqnarray}
where $f_{P,j_\ell}$ denotes the decay constant defined in eq.
(\ref{decay}). $\bar\Lambda_{P,j_\ell}=m_{P,j_\ell}-m_Q$.

We calculate the correlator (\ref{c1}) on the light-cone to the
leading order of ${\cal O}(1/ m_Q)$. The expression for $G_1(\omega,
\omega')$ reads
\begin{eqnarray}\label{f1}
&&-{\sqrt{6}\over 8}i \int_0^{\infty} dt \int dx e^{ikx} \delta
(x-vt){\rm Tr} \Big[ (\mathcal{D}^t_\alpha -{1\over
3}\gamma^t_\alpha {
\mathcal{D}\!\!\!\slash}^t )\nonumber\\
&&\qquad\qquad\qquad \times (1+{v\!\!\!\slash})\gamma_5 \langle \pi
(q)|u(0) {\bar d}(x) |0\rangle \Big]\; .
\end{eqnarray}

The pion (or $K$/$\eta$) distribution amplitudes are defined as the
matrix elements of nonlocal operators between the vacuum and pion
state. Up to twist four they are \cite{ball,ball2}:
\begin{eqnarray}
&&\langle\pi(q)| {\bar d} (x) \gamma_{\mu} \gamma_5 u(0)
|0\rangle=-i f_{\pi} q_{\mu} \int_0^1 du \;
e^{iuqx}\Big[\varphi_{\pi}(u)\nonumber\\
&&\quad +\frac{1}{16}m_{\pi}^2x^2 A(u)\Big]-\frac{i}{2} f_\pi
m_{\pi}^2 {x_\mu\over q x}
\int_0^1 du \; e^{iuqx}  B(u)\;,\nonumber\\
&&\langle\pi(q)| {\bar d} (x) i \gamma_5 u(0) |0\rangle = f_{\pi}
\mu_{\pi}
 \int_0^1 du \; e^{iuqx} \varphi_P(u)\;, \nonumber\\
&&\langle\pi(q)| {\bar d} (x) \sigma_{\mu \nu} \gamma_5 u(0)
|0\rangle
={i\over 6}(q_\mu x_\nu-q_\nu x_\mu) f_{\pi} \mu_{\pi}\nonumber\\
&&\qquad\qquad\qquad\qquad\times \int_0^1 du \; e^{iuqx}
\varphi_\sigma(u)\;.
\end{eqnarray}
The expressions for the light cone wave functions
$\varphi_{\pi}(u)$ etc are presented in Appendix \ref{appendix2}
together with the relevant parameters for $\pi$, $K$ and $\eta$.

Expressing eq. (\ref{f1}) with the light cone wave functions, we get
the expression for the correlator function in the quark-gluon level
\begin{eqnarray}\label{q1}
&&G_1(\omega, \omega')= -i{\sqrt{6}\over 12}f_\pi\int_0^{\infty}
dt \int_0^1 du e^{i (1-u) {\omega t \over 2}} e^{i u {\omega' t
\over 2}} u \nonumber\\
&&\quad\times\Big \{ {i \over t}[u\varphi_{\pi} (u)]'+{1\over
16}m_{\pi}^2[uA(u)]'+{1\over 2}B(u)\Big[{iu\over q\cdot v}\nonumber\\
&&\quad -{1\over (q\cdot v)^2t}\Big]+\mu_{\pi}u\varphi_p(u)+{1 \over
6}\mu_{\pi}\varphi_{\sigma}(u) \Big \} +\cdots\;.
\end{eqnarray}

After performing wick rotation and double Borel transformation with
the variables $\omega$ and $\omega'$ the single-pole terms in eq.
(\ref{pole}) are eliminated and we arrive at the following result:
\begin{eqnarray}\label{g1}
&&g_1 f_{-,{1\over 2} } f_{-, {3\over 2}
}\nonumber\\&&={\sqrt{6}\over 6}f_{\pi} \exp\Big[{ {
\Lambda_{-,{1\over 2} } +\Lambda_{-,{3\over 2} } \over T }}\Big]
\bigg\{ \frac{1}{2}[u\varphi_{\pi} (u)]^{'}T^2
f_1\Big({\omega_c\over T}\Big)\nonumber\\
&&-\frac{1}{8}m_{\pi}^2[uA(u)]^{'}+m_{\pi}^2[G_1(u)+G_2(u)]\nonumber\\
&&-\mu_{\pi} [u\varphi_P (u)+\frac{1}{6}\varphi_{\sigma}(u)]T
f_0\Big({\omega_c\over T}\Big)\bigg\}\bigg|_{u=u_0}\;,
\end{eqnarray}
where $u_0={T_1\over T_1+T_2}$, $T={T_1T_2\over T_1+T_2}$. $T_1$,
$T_2$ are the Borel parameters,  and
$f_n(x)=1-e^{-x}\sum\limits_{k=0}^{n}{x^k\over k!}$. The factor
$f_n$ is used to subtract the integral $\int_{\omega_c}^\infty s^n
e^{-{s\over T}} ds$ as a contribution of the continuum. The sum
rules we have obtained from the correlators (\ref{c1})-(\ref{c9})
are collected in Appendix \ref{appendix3} together with the
definitions of $G_1$ etc.

\section{Numerical results}\label{numerical}

For the ground states and P wave heavy mesons, we will use
\cite{braun,slz}:
\begin{eqnarray}
\label{fvalue} &&\bar\Lambda_{-,{1\over2}}=0.5
~\mbox{GeV}\;,\hspace{0.8cm} f_{-,{1\over2}}=0.25 ~\mbox{GeV}^{3/2}\;,\nonumber\\
&&\bar\Lambda_{+,{1\over2}}=0.85 ~\mbox{GeV}\;,\hspace{0.6cm}
f_{+,{1\over2}}=0.36\pm 0.10
~\mbox{GeV}^{1/2}\;,\nonumber\\
&&\bar\Lambda_{+,{3\over2}}=0.95 ~\mbox{GeV}\;,\hspace{0.6cm}
f_{+,{3\over2}}=0.26\pm 0.06 ~\mbox{GeV}^{5/2}\;.\nonumber
\end{eqnarray}
The mass parameters and decay constants for the D wave doublets have
been obtained in section \ref{mass} from the two point sum rule:
\begin{eqnarray}
\label{mass constant} &&\bar\Lambda_{-,{3\over2}}=1.42
~\mbox{GeV}\;,\hspace{0.2cm}f_{-,{3\over2}}=0.39\pm 0.03 ~\mbox{GeV}^{5/2}\;,\nonumber\\
&&\bar\Lambda_{-,{5\over2}}=1.38
~\mbox{GeV}\;,\hspace{0.2cm}f_{-,{5\over2}}=0.33\pm 0.04
~\mbox{GeV}^{7/2}\;.\nonumber
\end{eqnarray}

We choose to work at the symmetric point $T_1 = T_2 = 2T$, i.e.,
$u_0 = 1/2$ as traditionally done in literature \cite{lc}. The
working region for $T$ can be obtained by requiring that the higher
twist contribution is less than $30\%$ and the continuum
contribution is less than $40 \%$ of the whole sum rule, we then get
$\omega_c=3.2-3.6$ GeV and the working region $2.0<T<2.5$ GeV for
eqs. (\ref{b1}), (\ref{b2}) and (\ref{b3}) in Appendix
\ref{appendix3} and $1.2<T<2.0$ GeV for others. The working regions
for the first three sum rules are higher than that for the others
because there are zero points between 1 and $2$ GeV for them and
stability develops only for $T$ above $2$ GeV. From eq. (\ref{g1})
the coupling reads
\begin{eqnarray}
&&g_{1\pi}f_{-,{1\over 2} } f_{-, {3\over 2} }
   =(0.17\pm 0.04)~\mbox{GeV}^{3}\;.
\end{eqnarray}
We use the central values for the mass parameters and the error is
due to the variation of $T$ and the uncertainty of $\omega_c$. The
central value corresponds to $T=1.6$ GeV and $\omega_c=3.4$ GeV.
There is cancelation between the twist 2 and twist 3 contributions
in the sum rule.

For D wave heavy mesons with a strange quark, the couplings can be
obtained by the same way. Notice that in the $\eta$ case, $f_{\pi}$
should be replaced by $-{2 \over \sqrt{6}}f_{\eta}$ due to the quark
components of $\eta$ meson, where $f_{\eta}=0.16$ GeV is the decay
constant of $\eta$ meson. From eq. (\ref{g1}) we can get the
couplings between the ground state doublet and D wave doublet with a
strange quark,
\begin{eqnarray}
&&g_{1K}f_{-,{1\over 2} } f_{-, {3\over 2} }
   =(0.19\pm 0.06)~\mbox{GeV}^{3}\;,\nonumber\\
&&g_{1\eta}f_{-,{1\over 2} } f_{-, {3\over 2} }
   =(0.28\pm 0.06)~\mbox{GeV}^{3}\;.
\end{eqnarray}

The couplings between the $\frac{3}{2}^-$ and $\frac{5}{2}^-$
doublets and other doublets are collected in Table\;\ref{table}. We
can see that the $SU(3)_f$ breaking effect is not very big here.

\begin{table}[htb]
\begin{center}
\begin{tabular}{c|ccccccccccc} \hline\hline
$\frac{3}{2}^-$&$\frac{1}{2}^-$ &$\frac{1}{2}^+$
&${\frac{3}{2}}^+_d$&${\frac{3}{2}}^+_s$ &${\frac{3}{2}}^-_f$&
${\frac{3}{2}}^-_p$
\\\hline $\pi$&0.17&0.086&0.16&0.10&0.056&0.071\\\hline
$K$&0.19&0.09&0.24&0.18&0.057&0.10\\\hline
$\eta$&0.28&0.046&0.22&0.11&0.030&0.078\\\hline\hline
$\frac{5}{2}^-$&$\frac{1}{2}^-$ &$\frac{1}{2}^+$
&$\frac{3}{2}^+$&${\frac{3}{2}}^-_f$& ${\frac{3}{2}}^-_p$&
${\frac{5}{2}}^-_g$&${\frac{5}{2}}^-_f$&${\frac{5}{2}}^-_p$
\\\hline $\pi$&0.11&0.36&0.072&0.13&0.12&0.015&0.05&0.01\\\hline
$K$&0.14&0.48&0.083&0.11&0.16&0.015&0.09&0.02\\\hline
$\eta$&0.12&0.42&0.074&0.10&0.14&0.008&0.08&0.01\\\hline\hline
\end{tabular}
\end{center}
\caption{\label{table}The pionic couplings between $\frac{3}{2}^-$
and $\frac{5}{2}^-$ doublets and other doublets. The values are
the product of coupling constants and the decay constants of
initial and final heavy mesons. }
\end{table}

With the  central values of $f$'s, we get the absolute values of the
coupling constants:
\begin{eqnarray}
 &&g_{1\pi}
   =(1.74\pm 0.43)~\mbox{GeV}^{-1}\;,\nonumber\\
 &&g_{1K}
   =(1.95\pm 0.63)~\mbox{GeV}^{-1}\;,\nonumber\\
 &&g_{1\eta}
   =(2.87\pm 0.65)~\mbox{GeV}^{-1}\;.
\end{eqnarray}
For the ${5 \over 2}^-$ doublet we have
\begin{eqnarray}
&&g_{5\pi}
   =(1.33\pm 0.29)~\mbox{GeV}^{-3}\;,\nonumber\\
 &&g_{5K}
   =(1.70\pm 0.42)~\mbox{GeV}^{-3}\;,\nonumber\\
 &&g_{5\eta}
   =(1.45\pm 0.30)~\mbox{GeV}^{-3}\;.
\end{eqnarray}
We do not include the uncertainties due to $f$'s here.

\begin{figure}[hbt]
\begin{center}
\scalebox{0.8}{\includegraphics{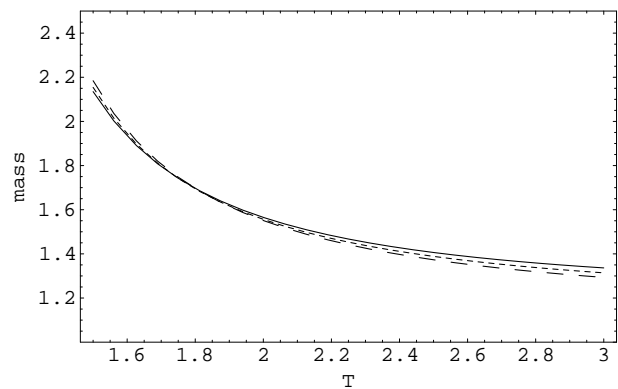}}
\end{center}
\caption{\label{fig3}The variation of the mass parameter
$\bar\Lambda_{-,{3\over2}}$ with the threshold $\omega_c$ (in unit
of $\mbox{GeV}$) for the ${3\over 2}^-$ doublet from LCQSR. The
long-dashed, short dashed and solid curves correspond to
$\omega_c$=3.2, 3.4, 3.6 GeV respectively.}
\end{figure}

We can also extract the mass parameter from the strong coupling
formulas obtained in the last section. By putting the exponential
factor on the left side of eq. (\ref{b2}) and differentiating it,
one obtains
\begin{equation}
\bar{\Lambda}_{-,{3\over 2} }={{T^2}\over 2} {{\rm
d}[\varphi_{\pi}(u_0)Tf_0({\omega_c\over
T})-\frac{1}{4}m_{\pi}^2A(u_0){1 \over T}]/{\rm d T} \over
[\varphi_{\pi}(u_0)Tf_0({\omega_c\over
T})-\frac{1}{4}m_{\pi}^2A(u_0){1 \over
T}-\frac{1}{3}\mu_{\pi}\varphi_{\sigma}(u_0)]}\;.
\end{equation}
With $\omega_c=3.2-3.6$ and the working region $2.0<T<2.5$ GeV, we
get
\begin{equation}
\bar\Lambda_{1,-,{3\over2}}=1.36-1.56 ~\mbox{GeV}\;,
\end{equation}
which is consistent with the value obtained by two point sum rule.
We present the variation of mass with $T$ and $\omega_c$ in
Fig.\;\ref{fig3}.

\section{Strong decay widths for D wave heavy mesons}\label{width}

Having calculated the coupling constants, one can obtain the pionic
decay widths of D wave heavy mesons. The widths for D wave states
decaying to $0^-$, $1^-$, $1^+$ states are
\begin{eqnarray}
&&\Gamma(B_{1}^{*'}\rightarrow
B^0\pi^-)=\frac{1}{24\pi}\frac{M_{B}}{M_{B_{1}^{*'}}}g_{1}^2|\bm{p}_1|^3\;,\nonumber\\
&&\Gamma(B_{1}^{*'}\rightarrow
B^{*0}\pi^-)=\frac{1}{12\pi}\frac{M_{B^{*}}}{M_{B_{1}^{*'}}}g_{1}^2|\bm{p}_1|^3\;,\nonumber\\
&&\Gamma(B_{1}^{*'}\rightarrow
B_1^0\pi^-)=\frac{1}{36\pi}\frac{M_{B_1}}{M_{B_{1}^{*'}}}g_{2}^2|\bm{p}_1|^5\;,\nonumber\\
&&\Gamma(B_{2}^{*}\rightarrow
B^{*0}\pi^-)=\frac{1}{36\pi}\frac{M_{B^{*}}}{M_{B_2}}g_{1}^2|\bm{p}_1|^3\;,\nonumber\\
&&\Gamma(B_{3}\rightarrow
B^0\pi^-)=\frac{1}{140\pi}\frac{M_{B}}{M_{B_{3}}}g_{5}^2|\bm{p}_1|^7\;,\nonumber\\
&&\Gamma(B_{3}\rightarrow
B^{*0}\pi^-)=\frac{1}{105\pi}\frac{M_{B^{*}}}{M_{B_{3}}}g_{5}^2|\bm{p}_1|^7\;,
\end{eqnarray}
where $|\bm{p}_1|$ is the moment of final state $\pi$. Note that
$g(B_2^{*}B^{*})=\sqrt{\frac{2}{3}}\;g({B_1^{*'}B^{*}})=\sqrt{\frac{2}{3}}\;g_1$.

\subsection{Nonstrange case}
We take 2.8 GeV and 6.2 GeV for the masses of D wave charmed and
bottomed mesons respectively. $M_D=1.87$ GeV, $M_{D^*}=2.01$ GeV,
$M_{D_1}=2.42$ GeV, $M_B=5.28$ GeV, $M_{B^*}=5.33$ GeV \cite{pdg}
and $M_{B_1}=5.75$ GeV from quark model prediction \cite{quark
model}. After summing over the charged and neutral modes, we get the
results listed in Table \ref{table2}.
\begin{table}[htb]
\begin{center}
\begin{tabular}{c|ccc|c|c} \hline\hline
&$D\pi$ &$D^{*}\pi$& $D_1\pi$&&$D^{*}\pi$
\\\hline $D_{1}^{*'}\rightarrow$&9-27&13-39&0.2&$D_{2}^{*}\rightarrow$&5-13\\\hline
\hline &$B\pi$&$B^{*}\pi$&$B_1\pi$&&$B^{*}\pi$
\\\hline $B_{1}^{*'}\rightarrow$&16-46&27-79&0.3&$B_{2}^{*}\rightarrow$&9-27\\\hline
\hline
\end{tabular}
\end{center}
\caption{\label{table2}The decay widths (in unit MeV) of the
charmed and bottomed D wave ($1^-, 2^-$) to ground doublets and
$\pi$.}
\end{table}

\subsection{Strange case}

We use $M_{D_s}=1.97$ GeV, $M_{D_s^{*}}=2.11$ GeV, $M_{B_s}=5.37$
GeV, $M_{B_s^{*}}=5.41$ GeV \cite{pdg}. Then for the charm-strange
sector we have Table \ref{table3}.
\begin{table}[htb]
\begin{center}
\begin{tabular}{c|ccccccccccc} \hline\hline
&$DK$ &$D^{*}K$ &$D_s\eta$&$D_s^{*}\eta$
\\\hline $D_{s1}^{*'}\rightarrow$&8-28&10-48&8-22&8-20\\\hline
$D_{s2}^{*}\rightarrow$&&4-16&&3-7\\\hline\hline &$BK$ &$B^{*}K$
&$B_s\eta$&$B_s^{*}\eta$
\\\hline $B_{s1}^{*'}\rightarrow$&12-52&18-84&11-27&16-42\\\hline
$B_{s2}^{*}\rightarrow$&&6-28&&6-14\\\hline\hline
\end{tabular}
\end{center}
\caption{\label{table3}The decay widths (in unit MeV) of the
charm-strange and bottom-strange D wave ($1^-, 2^-$) to ground
doublets and $K$/$\eta$.}
\end{table}
We do not consider the $D K^{*}$ mode and three-body modes in the
present work.

For the $2^-$, $3^-$ states with $j_{\ell}={5\over2}$, we find that
the widths are quite small and the branching fraction is perhaps
more useful. In the charm-strange sector the ratio of widths
(central values) for $DK$, $D^{*}K$, $D_s\eta$ and $D^{*}\eta$ modes
is $1:0.4:0.1:0.02$.

\section{Conclusion}\label{summary}

In this work we extract the mass and decay constants using the
traditional two point sum rule and calculate the strong couplings of
D wave heavy meson doublets with light hadrons $\pi$, $K$ and $\eta$
using LCQSR in the leading order of HQET. We also extract the mass
parameter from LCQSR for the coupling within the same D wave
doublet. The extracted mass parameters from two approaches are
consistent with each other. We then calculate the widths of D wave
heavy mesons decaying to light hadrons.

We have not considered the $1/m_{Q}$ correction and radiative
corrections. Heavy quark expansion works well for $B$ mesons where
the $1/m_b$ correction is under control and not so large. However,
the $1/m_{c}$ correction is not so small for the charmed mesons. It
will be desirable to consider both the $1/m_{Q}$ and radiative
corrections in the future investigation.

According to our present calculation, the ratios such as
${\Gamma(D_{sJ}(2860)\rightarrow DK)\over
\Gamma(D_{sJ}(2860)\rightarrow D_s\eta)}$ are useful in
distinguishing various interpretations of $D_{sJ}(2860)$ and
$D_{sJ}(2715)$. Treating $D_{sJ}(2860)$ as a D wave $1^-$ state, we
find the above ratio is $0.4-2.2$. If it is the radial excitation of
$D_s^\ast$, this ratio is 0.09 \cite{bozhang}.

The pionic widths of D wave states are not very large. With a mass
of $2.86$ GeV, the partial decay width of the $1^-$ D wave $D_s$
state into $DK$ and $D\eta$ modes is $34-118$ MeV. With a mass of
2.715 GeV its pionic width is $15-57$ MeV. Note that $DK^\ast$ modes
may be equally important. So detection of other decay channels, such
as $D_s\eta$ and $D^{*}K$ modes, will be very helpful in the
classification of these new states.

{\bf Acknowledgments:} W. Wei thanks P. Z. Huang for discussions.
This project is supported by the National Natural Science Foundation
of China under Grants 10375003, 10421503 and 10625521, Ministry of
Education of China, FANEDD and Key Grant Project of Chinese Ministry
of Education (NO 305001). X.L. thanks the support from the China
Postdoctoral Science Foundation (NO 20060400376).

\appendix

\section{}\label{appendix1}
The $T^{\alpha,\beta;\;\mu,\nu}$ and
$T^{\alpha,\beta,\lambda;\;\mu,\nu,\sigma}$ are defined as
\begin{eqnarray}
T^{\alpha,\beta;\;\mu,\nu}&=&\frac{1}{2}(g^{\alpha\mu}g^{\beta\nu}+g^{\alpha,\nu}g^{\beta,\mu})
-\frac{1}{3}g_{t}^{\alpha\beta}g_{t}^{\mu\nu} \; ,\label{ax} \nonumber\\
T^{\alpha,\beta,\lambda;\;\mu,\nu,\sigma}&=&
\frac{1}{6}(g^{\alpha\mu}g^{\beta\nu}g^{\lambda\sigma}+g^{\alpha\mu}g^{\beta\sigma}g^{\lambda\beta}
+g^{\alpha\nu}g^{\beta\mu}g^{\lambda\sigma}\nonumber\\
&+&g^{\alpha\nu}g^{\beta\sigma}g^{\lambda\mu}+g^{\alpha\sigma}g^{\beta\mu}g^{\lambda\nu}
+g^{\alpha\sigma}g^{\beta\nu}g^{\lambda\mu})\nonumber\\
&-&\frac{1}{9}(g_{t}^{\alpha\beta}g_{t}^{\mu\nu}g_{t}^{\lambda\sigma}
+g_{t}^{\alpha\lambda}g_{t}^{\mu\nu}g_{t}^{\beta\sigma}+g_{t}^{\beta\lambda}g_{t}^{\mu\nu}g_{t}^{\alpha\sigma})\;.\nonumber
\label{pscal}
\end{eqnarray}
The tensor structures for G wave, F wave and two P wave decays in
eq. (\ref{c9}) for the coupling between the two $\frac{5}{2}^{-}$
state are
\begin{eqnarray}
&&T^g=T^{\mu,\nu,\sigma;
\mu_1,\nu_1,\sigma_1}T^{\alpha,\beta;\alpha_1,\beta_1}q_{\mu_1}q_{\nu_1}q_{\sigma_1}q_{\alpha_1}q_{\beta_1}\;,\nonumber\\
&&T^{f}=\frac{1}{3}\Big\{\Big[q_t^{\mu}q_t^{\alpha}q_t^{\beta}-
 \frac{1}{6}q^2_t(g_t^{\mu\alpha}q_t^{\beta} +
 g_t^{\mu\beta}q_t^{\alpha} \nonumber\\
&&\qquad + {4\over
 3}g_t^{\alpha\beta}q_t^{\mu})\Big]g_t^{\nu\sigma}+(\mu,\nu,\sigma)\Big\}\;,\nonumber\\
&&T^{p1}=\frac{1}{6}\Big[g_t^{\mu\alpha}g_t^{\nu\sigma}
q_t^{\beta}+g_t^{\mu\beta}g_t^{\nu\sigma}q_t^{\alpha}+(\mu,\nu,\sigma)\Big]\;,\nonumber\\
&&
T^{p2}=\frac{1}{3}(q_t^{\mu}g_t^{\nu\sigma}+q_t^{\nu}g_t^{\mu\sigma}+q_t^{\sigma}g_t^{\mu\nu})g_t^{\alpha\beta}\;.\nonumber
\end{eqnarray}

\section{}\label{appendix2}


The distribution amplitudes $\varphi_{\pi}$ etc can be
parameterized as \cite{ball,ball2}
\begin{eqnarray}
\varphi_{\pi}(u) &=&
6u\bar{u}\bigg[1+a_1C^{3/2}_1(\zeta)+a_2C^{3/2}_2(\zeta)\bigg]\; ,\nonumber\\
\phi_p(u) &=& 1+\bigg[30\eta_3-\frac 52 \rho_{\eta}^2\bigg]
C^{1/2}_2(\zeta)+\bigg[-3\eta_3\omega_3\nonumber\\&&-\frac{27}{20}\rho_{\eta}^2-\frac
{81}{10}\rho_{\eta}^2a_2\bigg] C^{1/2}_4(\zeta)  \; ,\nonumber
\end{eqnarray}
\begin{eqnarray}
\phi_{\sigma}(u)&=&
6u(1-u)\bigg\{1+\bigg(5\eta_3-\frac{1}{2}\eta_3\omega_3-\frac{27}{20}\rho_{\eta}^2\nonumber\\
&&-\frac {3}{5}\rho_{\eta}^2a_2\bigg)\bigg\} C^{3/2}_2(\zeta)  \;
,\nonumber\\
g_{\pi}(u)&=&
1+\Big[1+\frac{18}{7}a_2+60\eta_3+\frac{20}{3}\eta_4\Big]
C^{1/2}_2(\zeta)\nonumber\\
&&+\Big[-\frac{9}{28}a_2-6\eta_3\omega_3\Big]C^{1/2}_4(\zeta)
\;,\nonumber
\end{eqnarray}
\begin{eqnarray}
A(u) &=& 6u\bar u \bigg\{ \frac{16}{15} +
\frac{24}{35}a_2 + 20 \eta_3 + \frac{20}{9}\eta_4 + \Big[-\frac{1}{15} + \frac{1}{16}\nonumber\\
&&-\frac{7}{27} \eta_3 \omega_3 - \frac{10}{27}\eta_4
\Big]C_2^{3/2}(\zeta) + \Big[ -\frac{11}{210} a_2 \nonumber\\&&-
\frac{4}{135}\eta_3\omega_3 \Big]C_4^{3/2}(\zeta) \bigg\} +
\Big(-\frac{18}{5} a_2 + 21\eta_4\omega_4 \Big)\nonumber\\&&\bigg\{
2 u^3 (10-15 u + 6 u^2) \ln u + 2\bar u^3 (10-15\bar u\nonumber\\&&
+ 6 \bar u^2)\ln\bar u + u \bar u (2 + 13u\bar u)\bigg\} \; ,
\end{eqnarray}
where $\bar{u} \equiv 1-u,\; \zeta \equiv 2u-1$.
$C^{3/2,1/2}_{1,2}(\zeta)$ are Gegenbauer polynomials. Here
$g_{\pi}(u)=B(u)+\varphi_{\pi}(u)$. $a_1^{\pi,\eta}=0,
a_1^{K}=0.06$, $a_2^{\pi,K,\eta}=0.25$, $\eta_3^{\pi,K}=0.015$,
$\eta_3^{\eta}=0.013$, $\omega_3^{\pi,K,\eta}=-3$,
$\eta_4^{\pi}=10$, $\eta_4^K=0.6$, $\eta_4^{\eta}=0.5$,
$\omega_4^{\pi, K, \eta}=0.2$. $\rho_{\pi}^2$ etc give the mass
corrections and are defined as $\rho_{\pi}^2=\frac
{(m_u+m_d)^2}{m_{\pi}^2}$, $\rho_{K}^2=\frac{m_s^2}{m_K^2}$,
$\rho_{\eta}^2=\frac {m_s^2}{m_{\eta}^2}$. $m_s=0.125$ GeV.
$\mu_{\pi}={m_{\pi}^2\over m_u+m_d}(1-\rho_{\pi}^2)$,
$\mu_{\tiny{K},\eta}={m_{\tiny{K},\eta}^2\over
m_s}(1-\rho_{K,\eta}^2)$. $f_{\pi}=0.13$ GeV, $f_{K}=0.16$ GeV,
$f_{\eta}=0.156$ GeV. All of them are scaled at $\mu=1~\mbox{GeV}$.

\section{}\label{appendix3}

In this appendix we collect the sum rules we have obtained for the
strong couplings of D wave heavy doublets with light hadrons.
\begin{widetext}
\begin{eqnarray}
g_1 f_{-,{1\over 2} } f_{-, {3\over 2} }&=&{\sqrt{6}\over
6}f_{\pi} \exp\bigg[ { \Lambda_{-,{1\over 2} }
+\Lambda_{-,{3\over 2} } \over T }\bigg] \bigg\{
\frac{1}{2}[u\varphi_{\pi} (u)]^{'}T^2 f_1\Big({\omega_c\over
T}\Big)
-\frac{1}{8}m_{\pi}^2[uA(u)]^{'}+m_{\pi}^2[G_1(u)+G_2(u)]\nonumber\\
 &&-\mu_{\pi} \Big[u\varphi_P (u)+\frac{1}{6}\varphi_{\sigma}(u)\Big]T
f_0\Big({\omega_c\over T}\Big)\bigg\}\bigg|_{u=u_0}\;,
\end{eqnarray}

\begin{eqnarray}
g_2 f_{+,{1\over 2} } f_{-, {3\over 2} }&=&{\sqrt{6}\over
4}f_{\pi} \exp\bigg[ { \Lambda_{+,{1\over 2} }
+\Lambda_{-,{3\over 2} } \over T }\bigg] u\bigg\{
\varphi_{\pi}(u)Tf_0\Big({\omega_c\over
T}\Big)-\frac{1}{4}m_{\pi}^2A(u)\frac{1}{T}
-\frac{1}{3}\mu_{\pi}\varphi_{\sigma}(u)\bigg\}\bigg|_{u=u_0}\;,\label{b1}
\end{eqnarray}

\begin{eqnarray}
 g_3^d f_{+,{3\over 2}}f_{-,{3\over 2} }&=&\frac{1}{8}f_{\pi}
\exp\bigg[ { \Lambda_{+,{3\over 2} } +\Lambda_{-,{3\over 2} }
\over T }\bigg] \bigg\{
[(u(1-u)\varphi_{\pi}(u)]^{'}T^2f_1\Big({\omega_c\over T}\Big)
-\frac{1}{4}m_{\pi}^2[(u(1-u)A(u)]^{'}\nonumber\\&&
+2m_{\pi}^2\big[G_3(u)+G_4(u)+2G_5(u)\big]
+2\mu_{\pi}\big[u(1-u)\varphi_{p}(u)+\frac{1}{6}\varphi_{\sigma}(u)\big]Tf_0\Big(
{\omega_c\over T}\Big)\bigg\}\bigg|_{u=u_0}\;,
\end{eqnarray}

\begin{eqnarray}
g_3^s f_{+,{3\over 2}}f_{-,{3\over 2} }&=&-\frac{1}{48}f_{\pi}
\exp\bigg[ { \Lambda_{+,{3\over 2} } +\Lambda_{-,{3\over 2} }
\over T }\bigg] \bigg\{
[(u(1-u)\varphi_{\pi}(u)]^{'''}T^4f_3\Big({\omega_c\over
T}\Big)+2\mu_{\pi}\bigg[\Big(u(1-u)\varphi_{p}(u)\Big)^{''}\nonumber\\
&&
+\frac{1}{6}\varphi_{\sigma}(u)^{''}\bigg]T^3f_2\Big({\omega_c\over
T}\Big)
-m_{\pi}^2\Big[4\big(u(1-u)\varphi_{\pi}(u)\big)^{'}+\frac{1}{4}\big(u(1-u)A(u)\big)^{'''}\nonumber\\
&& +\frac{3}{2}A(u)^{'}-2(1-2u)B(u)
-2\big(u(1-u)B(u)\big)^{'}-4G_6(u)\Big]T^2f_1\Big({\omega_c\over
T}\Big)\;\nonumber\\&&
-8m_{\pi}^2\mu_{\pi}\;\Big[u(1-u)\varphi_{p}(u)+\frac{1}{6}\varphi_{\sigma}(u)\Big]\;Tf_0\Big({\omega_c\over
T}\Big)\bigg\}\bigg|_{u=u_0}\;,
\end{eqnarray}

\begin{eqnarray}
g_4^p f_{-,{3\over 2}}f_{-,{3\over 2}}&=&{\sqrt{6}\over
96}f_{\pi} \exp\bigg[{ { \Lambda_{-,{3\over 2} }
+\Lambda_{-,{3\over 2} }\over T }}\bigg]
\bigg\{m_{\pi}^2A(u)Tf_0\Big({\omega_c\over
T}\Big)-\frac{2}{3}\mu_{\pi}\big[(1-u)\varphi_{\sigma}(u)\big]^{'}T^2f_1\Big({\omega_c\over
T}\Big)\bigg\}\bigg|_{u=u_0}\;,
\end{eqnarray}

\begin{eqnarray}
g_4^f f_{-,{3\over 2}}f_{-,{3\over 2}}&=&{\sqrt{6}\over
4}f_{\pi} \exp\bigg[{ { \Lambda_{-,{3\over 2} }
+\Lambda_{-,{3\over 2} } \over T }}\bigg]
u(1-u)\bigg\{\varphi_{\pi}(u)Tf_0\Big({\omega_c\over
T}\Big)-\frac{1}{4T}m_{\pi}^2A(u)-\frac{1}{3}\mu_{\pi}\varphi_{\sigma}(u)
\bigg\}\bigg|_{u=u_0}\label{b2}\;,
\end{eqnarray}

\begin{eqnarray}
g_5 f_{-,{1\over 2} } f_{-, {5\over 2} }&=&{1\over 2}f_{\pi}
\exp\bigg[{ { \Lambda_{-,{3\over 2} } +\Lambda_{-,{3\over 2} } \over
T }} \bigg]u^2\bigg\{\varphi_{\pi}(u)Tf_0\Big({\omega_c\over
T}\Big)-\frac{1}{4T}m_{\pi}^2A(u)+\frac{1}{3}\mu_{\pi}\varphi_{\sigma}(u)
\bigg\}\bigg|_{u=u_0}\;,\end{eqnarray}

\begin{eqnarray}
g_6 f_{+,{1\over 2}}f_{-,{5\over 2} }&=&-{\sqrt{15}\over 20}f_{\pi}
\exp\bigg[{ { \Lambda_{+,{1\over 2} } +\Lambda_{-,{5\over 2} } \over
T }}\bigg]\bigg \{
\big[u^2\varphi_{\pi}(u)\big]^{'}T^2f_1\Big({\omega_c\over
T}\Big)-\frac{1}{4}m_{\pi}^2\big[u^2A(u)\big]^{'}\nonumber\\ &&
-\frac{1}{2}m_{\pi}^2\big[G_7(u)+2G_8(u)+2G_5(u)\big]
+2\mu_{\pi}u\Big[u\varphi_{p}(u)+\frac{1}{3}\varphi_{\sigma}(u)\Big]Tf_0\Big({\omega_c\over
T}\Big)\bigg\}\bigg|_{u=u_0}\;,\end{eqnarray}

\begin{eqnarray}
g_7 f_{+,{3\over 2}}f_{-,{5\over 2}}&=&-{\sqrt{10}\over 30}f_{\pi}
\exp\Big[{ { \Lambda_{+,{3\over 2} } +\Lambda_{-,{5\over 2} } \over
T }}\Big]
\bigg\{-\frac{1}{4}\big[u^2(1-u)\varphi_{\pi}(u)\big]^{''}T^3f_2\Big({\omega_c\over
T}\Big)-\frac{1}{12}\mu_{\pi}\Big[7\big(u(1-\frac{2}{7}u)\varphi_{\sigma}(u)\big)^{'}\nonumber\\&&
+\big(u^2(1-u)\varphi_{\sigma}(u)\big)^{''}\Big]T^2f_1\Big({\omega_c\over
T}\Big) +m_{\pi}^2\Big[u^2(1-u)\varphi_{\pi}(u)+\frac{7}{8}u
A(u)\nonumber\\&& +\frac{1}{16}\big(u^2(1-u)A(u)\big)^{''}
\Big]Tf_0\Big({\omega_c\over T}\Big)
+\frac{1}{3}m_{\pi}^2\mu_{\pi}u^2(1-u)\varphi_{\sigma}(u)
\;\bigg\}\bigg|_{u=u_0}\;,\end{eqnarray}

\begin{eqnarray}
g_8^p f_{-,{3\over 2}}f_{-,{5\over 2} }&=&{\sqrt{10}\over 18}f_{\pi}
\exp\Big[{ { \Lambda_{-,{3\over 2} } +\Lambda_{-,{5\over 2} } \over
T }}\Big]
\bigg\{\frac{1}{8}\big[u^2(1-u)\varphi_{\pi}(u)\big]^{'''}T^4f_3\Big({\omega_c\over
T}\Big)+\frac{1}{4}\mu_{\pi}\Big[\big(u^2(1-u)\varphi_{p}(u)\big)^{''}\nonumber\\&&
+\frac{1}{3}\big(u(1-\frac{5}{2}u)\varphi_{\sigma}(u)\big)^{''}\Big]T^3f_2\Big({\omega_c\over
T}\Big)-\frac{1}{2}m_{\pi}^2\Big[\big(u^2(1-u)\varphi_{\pi}(u)\big)^{'}+\frac{27}{40}\big(u
A(u)\big)^{'}\nonumber\\&&
+\frac{1}{16}\big(u^2(1-u)A(u)\big)^{'''}-\frac{1}{5}u\big(1-\frac{3}{2}u\big)
B(u)-\frac{1}{2}\big(u^2(1-u)B(u)\big)^{'} -G_9(u)\nonumber\\&&
+\frac{9}{5}G_{2}(u)\Big]T^2f_1\Big({\omega_c\over T}\Big)
-m_{\pi}^2\mu_{\pi}\Big[u^2(1-u)\varphi_{p}(u)+\frac{1}{3}\big(1-\frac{5}{2}u\big)\varphi_{\sigma}(u)\Big]Tf_0\Big({\omega_c\over
T}\Big)\bigg\}\bigg|_{u=u_0} \;,\end{eqnarray}

\begin{eqnarray}
g^f_8 f_{-,{3\over 2}}f_{-,{5\over 2} }&=&{\sqrt{10}\over 15}f_{\pi}
\exp\Big[{ { \Lambda_{1,-,{3\over 2} } +\Lambda_{2,-,{5\over 2}
}\over T }}\Big]
\bigg\{-\frac{1}{2}\big[u^2(1-u)\varphi_{\pi}(u)\big]^{'}T^2f_1\Big({\omega_c\over
T}\Big)+\frac{1}{8}m_{\pi}^2\big[u^2(1-u)A(u)\big]^{'}\nonumber\\ &&
+m_{\pi}^2\big[G_{10}(u)-2G_{11}(u)+2G_{12}(u)-6G_{13}(u)\big]\nonumber\\
&&
-\mu_{\pi}u(1-u)\Big[u\varphi_{p}(u)+\frac{1}{12}\varphi_{\sigma}(u)\Big]Tf_0\Big({\omega_c\over
T}\Big)\bigg\}\bigg|_{u=u_0}\;,\end{eqnarray}

\begin{eqnarray}
g_9^g f_{-,{5\over 2}}^{2}&=&{\sqrt{15}\over 6}f_{\pi} \exp\Big[{ {
\Lambda_{-,{5\over 2} } +\Lambda_{-,{5\over 2} } \over T }}\Big]
u^2(1-u)^2\bigg\{\varphi_{\pi}(u)Tf_0\Big({\omega_c\over
T}\Big)-\frac{1}{4}m_{\pi}^2A(u){1 \over
T}-\frac{1}{3}\mu_{\pi}\varphi_{\sigma}(u)
\bigg\}\bigg|_{u=u_0}\;,\nonumber\\\label{b3}\end{eqnarray}

\begin{eqnarray}
g_9^f f_{-,{5\over 2}}^{2}&=&{\sqrt{15}\over 45}f_{\pi} \exp\Big[{ {
\Lambda_{-,{5\over 2} } +\Lambda_{-,{5\over 2} } \over T }}\Big]
\bigg\{-\frac{1}{4}\big[u^2(1-u)^2\varphi_{\pi}(u)\big]^{''}T^3f_2\Big({\omega_c\over
T}\Big)+\frac{1}{12}{\mu}_{\pi}\Big[\big(u^2(1-u)^2\varphi_{\sigma}(u)\big)^{''}\;,\nonumber\\&&
-\frac{3}{8}\big(u(1-u)^2\varphi_{\sigma}(u)\big)^{'}\Big]T^2f_1\Big({\omega_c\over
T}\Big)+m_{\pi}^2\Big[u^2(1-u)^2\varphi_{\pi}(u)-\frac{1}{8}u(3+7u)A(u)\nonumber\\&&
-3\big(G_{10}(u)+2G_{11}(u)+2G_{12}(u)-6G_{13}(u)\big)\Big]Tf_0\Big({\omega_c\over
T}\Big)
-\frac{1}{3}m_{\pi}^2\mu_{\pi}u^2(1-u)^2\varphi_{\sigma}(u)\bigg\}\bigg|_{u=u_0}
,\end{eqnarray}

\begin{eqnarray}
g_9^{p1}f_{-,{5\over 2}}^{2}&=&-{\sqrt{15}\over 45}f_{\pi}
\exp\Big[{ { \Lambda_{2,-,{5\over 2} } +\Lambda_{3,-,{5\over 2} }
\over T }}\Big]
\bigg\{\frac{1}{24}\mu_{\pi}\big[u^2(1-u)\varphi_{\sigma}(u)\big]^{'''}T^4f_3\Big({\omega_c\over
T}\Big)\nonumber\\&&
-\frac{1}{8}m_{\pi}^2\Big[\frac{3}{2}(u(1-2u)A(u))^{''}+u(2-3u)B(u)+\big(u^2(1-u)B(u)\big)^{'}\nonumber\\&&
+2G_{9}(u)-6G_2(u)\Big]T^3f_2\Big({\omega_c\over T}\Big)
-\frac{1}{6}m_{\pi}^2\mu_{\pi}\big[u^2(1-u)^2\varphi_{\sigma}(u)\big]^{'}T^2f_1\Big({\omega_c\over
T}\Big)\bigg\}\bigg|_{u=u_0}\;,
\end{eqnarray}

The $G$'s are defined as integrals of light cone wave function
$B(u)$
\begin{eqnarray}
&&G_1 (u)\equiv \int_0^{u} t B(t)dt,\hspace{0.4cm} G_2 (u)\equiv \int_0^{u}dx\int_0^xB(t)dt\;,\nonumber\\
&&G_3 (u)\equiv \int_0^{u} t(1-t)B(t)dt,\hspace{0.4cm} G_4 (u)\equiv \int_0^{u}dx\int_0^x(1-2t)B(t)dt\;,\nonumber\\
&&G_5 (u)\equiv \int_0^{u}dx\int_0^xdy\int_0^y B(t)dt\;, \hspace{0.4cm} G_6 (u)\equiv \int_0^{u} B(t)dt\;,\nonumber\\
&&G_7 (u)\equiv \int_0^{u} t^2 B(t)dt\;, \hspace{0.4cm}G_8 (u)\equiv \int_0^{u}dx\int_0^x t B(t)dt\;,\nonumber\\
&&G_{9} (u)\equiv \int_0^{u} (1-3t) B(t)dt\;,\hspace{0.4cm}G_{10} (u)\equiv \int_0^{u} t^2(1-t) B(t)dt\;,\nonumber\\
&&G_{11} (u)\equiv \int_0^{u}dx\int_0^x t(1-\frac{3}{2}t)B(t)dt\;,\hspace{0.4cm}G_{12} (u)\equiv \int_0^{u}dx\int_0^xdy\int_0^y (1-3t)B(t)dt\;,\nonumber\\
&&G_{13} (u)\equiv \int_0^{u}dx\int_0^xdy\int_0^ydz\int_0^z
B(t)dt\;.\nonumber
\end{eqnarray}
\end{widetext}

\end{document}